\documentclass{elsart}
\usepackage{graphics,graphicx,amssymb}

\newbox\grsign \setbox\grsign=\hbox{$>$} \newdimen\grdimen \grdimen=\ht\grsign
\newbox\simlessbox \newbox\simgreatbox
\setbox\simgreatbox=\hbox{\raise.5ex\hbox{$>$}\llap
     {\lower.5ex\hbox{$\sim$}}}\ht1=\grdimen\dp1=0pt
\setbox\simlessbox=\hbox{\raise.5ex\hbox{$<$}\llap
     {\lower.5ex\hbox{$\sim$}}}\ht2=\grdimen\dp2=0pt

\begin{document}

\begin{frontmatter}

\title{Microeconomics of the ideal gas like market models}

\large{Anindya S. Chakrabarti$~^{1*}$, Bikas K. Chakrabarti$~^{1,2\dag}$}

\textit{\small{$~^1$ Economic Research Unit, Indian Statistical Institute, \\203 B. T. Road, Kolkata 700018, India}}

\textit{\small{$~^2$ Theoretical Condensed Matter Physics Division and Center for Applied Mathematics and Computational Science,\\
Saha Institute of Nuclear Physics,\\
1/AF Bidhannagar, Kolkata 700064, India}}

\thanks{Email addresses: $~^*$aschakrabarti@gmail.com (corresponding author), $~^\dag$bikask.chakrabarti@saha.ac.in}

\begin{abstract}
\noindent We develop a framework based on microeconomic theory from which
the ideal gas like market models can be addressed.
A kinetic exchange model based on that framework is proposed and 
its distributional features have been studied by considering
its moments. 
Next, we derive the moments of the CC model (Eur. Phys. J. B 17 (2000) 167) as well. 
Some precise solutions are obtained which
conform with the solutions obtained earlier. Finally, an output market
is introduced with global price determination
in the model with some necessary modifications. 
\end{abstract}
\end{frontmatter}

\section{Introduction}
{
\noindent Starting with an early attempt by Angle \cite{angle,anglephys}, a number
of models based on kinetic theory of gas
have been proposed to understand
the emergence of the universal features of income and wealth 
distributions (see e.g. refs. \cite{ccm,   chatterjee, rosser}). The main
focus of those models was to develop a framework that would give rise to
gamma function-like behavior for the bulk of the distribution and a power-law
for the richer section of the population. 
The CC-CCM models \cite{anirbanc, manna} have both of these features. The
kinetic exchange model proposed by Dragulescu and Yakovenko \cite{yakovenko} 
and later studied in
more details by Guala \cite{guala}, produces the gamma function-like
behavior for the income distribution. 
We note that all of these models
are generally based on some ad-hoc stochastic
asset evolution equations with little theoretical foundations for it. Our primary
aim in this paper is to develop a consistent framework from which we can
address this type of market models. Here we
propose a model based on consumers' optimization which can give
rise to those particular forms of asset exchange equations used in refs.
\cite{anirbanc, yakovenko} as special cases.
We then focus exclusively on the asset exchange equations
and an analytically simple kinetic exchange model is proposed.
Its distributional features are analyzed by considering its moments. 
The same technique is then applied to
derive the moments of the distribution of income in the CC model \cite{anirbanc} 
as well. We find
that it provides a rigorous justification for the values of the parameters of the 
distribution,
conjectured earlier in ref. \cite{kaski}. A possible extension of the microeconomic settings
of the basic model
is also studied where we consider the output market explicitly with global price
determination.
}

\section{The model}
{
\noindent We consider an $N$-agent exchange economy. Each of them produces a single 
perishable commodity. Each of these goods is different from all other goods. Money exists
in this economy to facilitate transactions (existence of money is not formally explained here). 
Any commodity can enter as an argument in the utility function (see ref. \cite{mwg} for
a detailed discussion on the {\it theory of utility}) of any agent.
These agents care for their future consumptions and hence they care
about their savings in the current period as well. Each of these agents
are endowed with an initial amount of money (the only type of non-perishable 
asset considered here) which is assumed to be unity
for every agent for simplicity.
At each time step, two agents meet randomly to carry out transactions according
to their utility maximization principle. We also assume that the agents have
time dependent preference structure. More precisely, we assume that
the parameters of the utility function can vary over time \cite{lux, jet}. 
In what follows,
we analyze the trading outcomes when any two such agents meet in the market at some 
time-step $t$.
 
\noindent Suppose agent 1 produces $Q_1$ amount of commodity 1 only and agent 2 produces
$Q_2$ amount of
commodity 2 only and the amounts of money in their possession at 
time $t$ are $m_1(t)$ and $m_2(t)$ respectively. Since neither of the two agents possess
the commodity produced by the other agent, both of them will be willing to trade and buy the 
other good by selling a fraction of their own productions
as well as with the money that they hold. In general, at each time step
there would be a net transfer
of money from one agent to the other due to trade.
Our aim is to understand how the amount of money held by the agents change over time.
For notational convenience, we denote $m_i(t+1)$ as $m_i$   
and $m_i(t)$ as $M_i$ (for $i=1, 2$).

We define the utility functions as follows.
For agent $1$, {$U_1(x_1,x_2,m_1)= x_1^{\alpha_1}x_2^{\alpha_2}m_1^{\alpha_m}$}
and for agent $2$,
$U_2(y_1,y_2,m_2)=y_1^{\alpha_1}y_2^{\alpha_2}m_2^{\alpha_m}$ where the
arguments in both of the utility functions are consumption of the first (i.e.
$x_1$ and $y_1$) and second good (i.e. $x_2$ and $y_2$)
and amount of money in their possession respectively.
For simplicity, we assume that the utility functions are of the Cobb-Douglas
form with the sum of the powers normalized to $1$ i.e. 
$\alpha_1+\alpha_2+\alpha_m=1$, which corresponds to the
constant returns to scale property (homogeneity of degree one)\cite{mwg}. 
Let the commodity prices  
to be determined in the market be denoted by $p_1$ and $p_2$.
Now, we can define the budget constraints as follows. For agent $1$
the budget constraint is $p_1x_1+p_2x_2+m_1\leq M_1+p_1Q_1$ and
similarly, for agent $2$ the constraint is
$p_1y_1+p_2y_2+m_2 \leq M_2+p_2Q_2$. 
What these constraints mean
is that the amount that agent $1$ can spend for consuming $x_1$ and $x_2$
added to the amount of money that he holds after trading at time $t+1$
(i.e. $m_1$) cannot exceed the amount of money that he has 
at time $t$ (i.e. $M_1$)
added to what he earns by selling the good he produces (i.e. $Q_1$). 
The same is true for
agent $2$. 

\noindent The basic idea behind this whole exercise is that both of the agents try to
maximize their respective utility subject to their respective budget constraints and
the {\it invisible hand} of the market that is the price mechanism 
works to clear the market for both goods (i.e. total demand equals
total supply for both goods at the equilibrium prices). Ultimately we will study the
money evolution equations in such a situation.  
Formally, agent 1's problem is to maximize his utility subject to
his budget constraint i.e. maximize $U_1(x_1,x_2,m_1)$
subject to $p_1.x_1+p_2.x_2+m_1=M_1+p_1.Q_1$.
Similarly for agent 2, the problem is to maximize $U_1(y_1,y_2,m_2)$
subject to $p_1.y_1+p_2.y_2+m_2=M_2+p_2.Q_2$.
Solving those two maximization exercises by Lagrange multiplier and applying the condition
that the market remains in equilibrium, we get the competitive
price vector ($\hat p_1, \hat p_2$) as 
$\hat p_i=(\alpha_i/\alpha_m)(M_1+M_2)/Q_i$ for $i=1$, $2$ (see appendix {\bf A1}).
 
We now examine the outcomes of such a trading process.
\begin{enumerate}

\item[(a)] 
At optimal prices $(\hat p_1, \hat p_2)$, $m_1(t)+m_2(t)=m_1(t+1)+m_2(t+1)$
and this follows directly from
Walras' law \cite{mwg} saying that if all but one market clears then the rest also
has to be cleared. That is, demand matches supply in all market at
the market-determined price in equilibrium. Since money is also
treated as a commodity in this framework, its demand (i.e. the total
amount of money held by the two persons after trade) must be equal to what
was supplied (i.e. the total amount of money held by them before trade).
In any case, an algebraic proof is also given in the appendix (see {\bf A2}).

\item[(b)] We now present the most important equation of money exchange in this model.
We make a rather restrictive assumption that $\alpha_1$ in the utility function 
can vary randomly over time with $\alpha_m$ remaining constant. It readily follows that $\alpha_2$ 
also varies randomly over time with the restriction that the sum of $\alpha_1$ and $\alpha_2$ is a
constant (1-$\alpha_m$).
In the money demand equations derived from the above-mentioned
problem, we substitute $\alpha_m$ by $\lambda$ and $\alpha_1/(\alpha_1+\alpha_2)$ 
by $\epsilon$ to get the following money evolution equations as (see {\bf A3})
$$m_1(t+1)=\lambda m_1(t)+\epsilon(1-\lambda)(m_1(t)+m_2(t))$$
$$m_2(t+1)=\lambda m_2(t)+(1-\epsilon)(1-\lambda)(m_1(t)+m_2(t)). \eqno(1)$$
\noindent For a fixed value of $\lambda$, if $\alpha_1$ (or $\alpha_2$, see {\bf A3}) 
is a random variable with uniform distribution over the domain $[0,1-\lambda]$,
then 
$\epsilon$ is also uniformly distributed over the domain $[0,1]$. 
It may be noted that $\lambda$ (i.e. $\alpha_m$ in the utility function) is 
the savings propensity used in the CC model \cite{anirbanc}.

\item[(c)] For the limiting value of $\alpha_m$ in the utility function
(i.e. $\alpha_m\rightarrow0$
which implies $\lambda\rightarrow0$), 
we get the money transfer
equation describing the random sharing of money without savings. This form of transfer
equation has been used in Dragulescu and Yakovenko \cite{yakovenko}, Guala
\cite{guala} and also
in the model proposed later in this paper.

\item[(d)] It may be noted that at each time step, the price mechanism 
works only locally ,i.e., it works
to clear the markets for two commodities (Q$_1$ and Q$_2$) only.
The markets considered here are perfectly competitive. 
Also, the set of competitive equilibria is a subset of the set of
Pareto Optimal allocation or in other words, all competitive
allocations are Pareto Optimal (see ref. \cite{mwg} for the definition of {\it Pareto
Optimality}). Hence, all the allocations achieved 
through such trading processes are Pareto Optimal. Also, since the exchange equations are not
sensitive to the level of
production, even if for some reason the
level of production alters (due to production shock) the form of the transfer
equations will remain the same
provided the form of the utility function remains the same.

\item[(e)] Ref \cite{jet} also presents a microeconomic framework alongwith an
asset evolution equation (see also \cite{lux}). 
But unlike here, the asset evolution equation for the $i-$th agent
in ref. \cite{jet} depends on his own assets only.

\end{enumerate}

\section{Stochastic model A: exchange with direct transfers}
{\noindent We now consider an economy under government supervision. Suppose that
the government believes in free market mechanism
and at the same time, it also wishes to restore equality by taxing the richer ones. In
other words, the government does not interfere with the process of market
transactions but wishes to restore equality by redistributing income
after each trading is performed.
There are, say $N$ agents (where $N$ is macroscopically large)
in the economy, each endowed with some amount
of money (which is normalized to unity, for simplicity)
at the begining of all tradings. At each time-point the money exchange process
takes place in two steps.

\noindent $(a)$ Two agents are randomly selected and they trade in an absolutely
random fashion. Note that this step directly followes from eqn. $(1)$ above if we
consider that $\lambda\rightarrow 0$ (i.e. if $\alpha_m$ in the utility function tends
to zero). Hence,
$$m_i(t+1/2)=\epsilon[m_i(t)+m_j(t)] $$
$$m_j(t+1/2)=(1-\epsilon)[m_i(t)+m_j(t)].     \eqno (2)$$
\noindent $(b)$ The agents agree to split the {\it excess} income. Hence the agent with more money, transfers a
fraction $f$ of the excess income to the agent with less money. It is reasonable to assume that
$0 \leq f \leq 0.5$.
\noindent If $m_i(t+1/2) \ge m_j(t+1/2)$,
excess income $\delta$ = $m_i(t+1/2)-m_j(t+1/2)$. Hence,
$$m_i(t+1)=m_i(t+1/2)-(f\delta ) $$
$$m_j(t)=m_j(t+1/2)+(f\delta ).     \eqno (2a)$$ 
\noindent This process is repeated at each time step until the
system reaches a steady state and the
distribution $p(m)$ of income among the agents in the steady state are studied.
The emergence of gamma function-like behavior is clearly seen in the figure [1].
It is seen computationally that as the {\it fraction} increases
from $0$ to $0.5$, the distribution becomes a delta function
starting from an exponential one. Left panel of figure [1] shows the pdf of income for several
values of $f$.
\begin{figure}
\begin{center}
\noindent \includegraphics[clip,width= 5cm,angle = 270]
{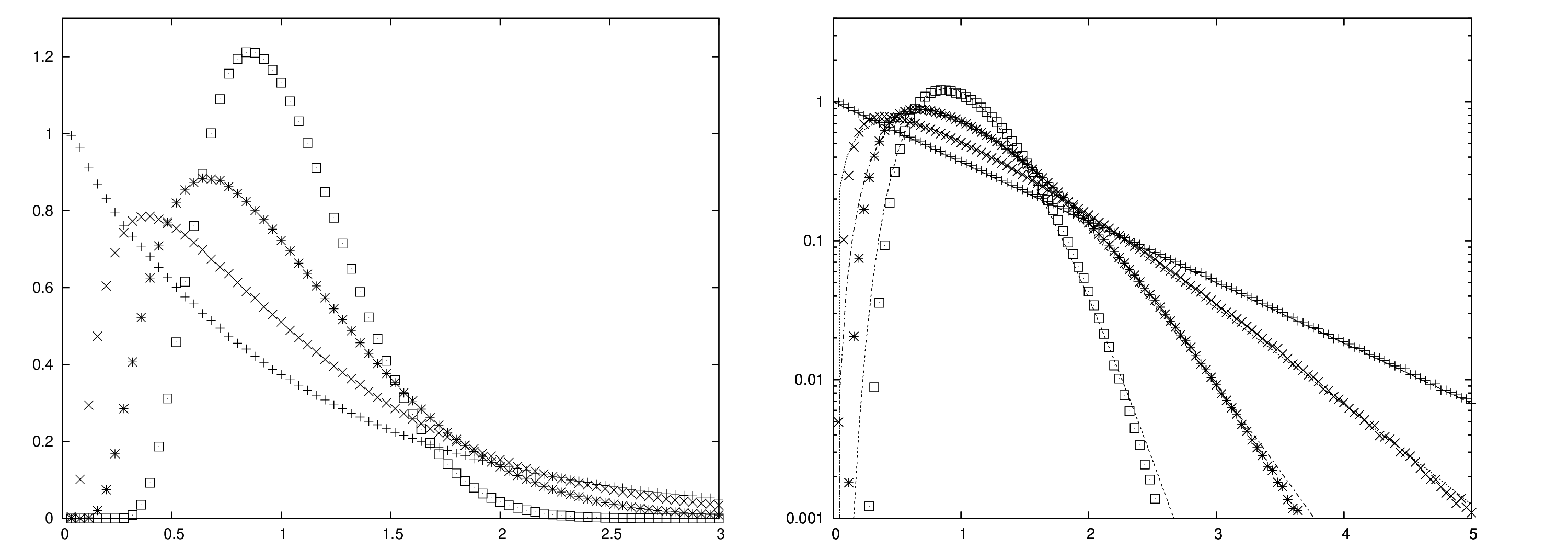}
\caption\protect{\label{fig:fig1}{\small MC results of transfer model with $N=100$. 
All simulations
are done for $10$ million time-steps and averaged over $1$ million time-steps.
+, $\times$, $\ *$ and $\square$ denotes the cases where $f$=0, 0.1,
0.2 and 0.3 respectively. {\it Left panel}: Steady
state income distributions of
the transfer model. {\it Right panel}: Steady state income distributions in semi-log scale with
gamma pdfs. (dotted lines).
}}
\end{center}
\end{figure}
\medskip

{\bf Qualitative features of the distribution}

\noindent Substituting for $\delta$, $m_i(t+1/2)$ and $m_j(t+1/2)$ in $(2a)$ 
we get the reduced equations
$$m_i(t+1)=g[m_i(t)+m_j(t)]$$
$$m_j(t+1)=(1-g)[m_i(t)+m_j(t)].     \eqno (3)$$ 
\noindent The expression of $g$ in the above equations is $g=f+(1-2f)\epsilon$.
It may be noted that $g$ is a linear transformation of an uniformly distributed
variable $\epsilon$.
Hence, $g$ is also uniformly distributed and its domain is $[f,1-f]$. 
Consider now the $i-$th agent only and analyze its income equation.
We denote {\it expectation} or average of a variable 
$x$ by $\langle x\rangle$ and variance of $x$
as $\Delta x$ where $\Delta x = \langle(x - \langle x\rangle)^2\rangle$.
By taking expectations on both sides of the income equation, it can be easily shown that
$\langle m \rangle =1 $ (see {\bf A4}). To proceed further we 
derive the following formula (see {\bf A4}) that relates variance of $g$ 
to that of $m$ i.e. the steady 
state money distribution,
$$\Delta m=\frac{4\Delta g}{\frac{1}{2}-2\Delta g}.  \eqno(4)$$
\noindent Also note that $g\sim uniform[f,1-f]$ where $0 \leq f \leq 0.5$ which implies
$\Delta g=(1-2f)^2/12$.
So when $f=0$, $\Delta g=1/12$ and this implies $\Delta m=1$ which is indeed the
variance of the distribution $p(m)=e^{-m}$. Again, if $f=0.5$, $\Delta g=0$ implying
$\Delta m=0$ which in turn implies the resulting distribution is a $\delta$ function.

{\bf Fit with Gamma distribution}

\noindent We fit the resulting distribution to a Gamma function
$$p(m)=\frac{m^{\alpha-1}e^{-\beta m}}{\Gamma(\alpha)\beta^{-\alpha}}.  \eqno(5)$$
It is well known that the first two moments of this distribution are $\alpha/\beta$
and $\alpha/\beta^2$ respectively. Comparing with what we
have found, we get $\alpha/\beta=1$ and
$1/\beta=4\Delta g/[\frac{1}{2}-2\Delta g]$. 
This fit is shown in the right panel of figure [1].

\section{Stochastic model B: exchange with savings}
\noindent We consider the following asset equations with a savings parameter $\lambda$
$$m_i(t+1)=\lambda m_i(t)+\epsilon(1-\lambda)[m_i(t)+m_j(t)]$$
$$m_j(t+1)=\lambda m_j(t)+(1-\epsilon)(1-\lambda)[m_i(t)+m_j(t)].   \eqno (6)$$
\noindent It may be noted that this is the equation used in the CC model \cite{anirbanc} and
has been derived by the utility maximization principle above (eqn. $(1)$). 
The CC model has been studied
extensively though the exact form of the distribution is still unknown. In ref. \cite{kaski},
it has been conjectured that the steady state distribution is approximately
a Gamma distribution [eqn. (5)] with parameter $\alpha=(1+2\lambda)/(1-\lambda)$.
Here, we give a simple derivation for this by fixing the
average amount of money per agent to unity.

As is done in the last model (section $2$), we consider the $i$-th agent only
$$m_i(t+1)=\lambda m_i(t)+\epsilon(1-\lambda)[m_i(t)+m_j(t)].$$
By taking expectation on both sides it may be shown that $\langle m \rangle = 1$ (see {\bf A5}).
Next, by applying the variance operator on both sides and simplifying, we get
$$\Delta m=\lambda^2(\Delta m+1)+2(1-\lambda)^2
(\Delta \epsilon+\frac{1}{4})(\Delta m+2)+\lambda(1-\lambda)
(\Delta m+2)-1.$$
Note that here $\epsilon\sim uniform[0,1]$ which implies $\Delta \epsilon=1/12$. Substituting
this value for $\Delta \epsilon$ and after rearranging terms we get,
$$\Delta m=\frac{(1-\lambda)^2}{(1-\lambda)(1+2\lambda)}.  \eqno(7)$$
The proof is given in appendix (see {\bf A5}).
Clearly if $\lambda \neq 1$, $\Delta m=(1-\lambda)/(1+2\lambda)$ as in ref. \cite{kaski}. 
Hence $\Delta m=1$ for $\lambda=0$, which is indeed the case for an exponential
distribution and for $0 \leq \lambda < 1$, the distribution is approximated
by eqn. (5) with $\alpha=(1+2\lambda)/(1-\lambda)$ and $\beta=\alpha$ as is
conjectured in ref. \cite{kaski}. 
For $\lambda\rightarrow1$, however, by applying l'H$\hat{o}$pitals rule
we get $\Delta m=0$. 
That explains why the steady state distribution tends to a delta 
function as the rate of savings i.e. $\lambda\rightarrow 1$ as widely observed in
simulations \cite{ccm, anirbanc, kaski, stinchcombe}.

\section{A generalized framework: trading in global market}
\noindent So far we have assumed a situation where at each period two agents meet and they
carry out transactions according to their utility considerations  . The price
mechanism works only locally, between these two agents to match supply and demand of
these two agents. We try to generalize the model to include an expanding market
where prices will be determined globally and the market will be cleared globally.
Here, we depict the market as a black box where agents 
interact with each other indirectly through the market. In the demand side,
each agent maximizes intertemporal utility subject to budget constraint and 
allocates income accordingly between present and future consumption. 
In the basic model (section $2$), present consumption
was expressed as a function of the two commodities consumed. But here we relax that
assumption and represent both present and future consumption by the amount of money spend on
present and future consumptions. What the agents save for future consumption
earns them interest income. Hence the market grows over time. 
On the production side, they can invest
in production and get their returns accordingly.    
Formally, there are $N$ number of agents in the economy each taking part
in the consumption and production activitites in each period.
Agents decide about future in discrete framework. The income stream
generated by an agent is also discrete. We assume that each agent when
deciding about the extent of present and future consumption, treats the problem as
a two-period choice problem with the present (today or this week or this month etc.)
as the first period and
everything after that as the second period. In each period, each agent
invests a part of his current money-holding to produce some goods and he gets
back some return from the market by selling them. Here we make another
assumption that the market is perfectly competitive which means that the number 
of agents is so large that an agent
by itself can not influence the pricing mechanism of the market.

\noindent We analyze a typical agent's behavior at any time step $t$ in the 
following three steps.
\begin{enumerate}
\item[(i)] Each agents problem is to maximize utility subject to
his budget constraint. For simplicity, we assume that the utility function
is of Cobb-Douglas type. Briefly, at time $t$ the $i-$th agent's problem is to
maximize $u(f,c)=f^{\lambda} c^{(1-\lambda)}$
subject to $f/(1+r)+c=m(t) $
where $f$ is the amount of money kept for future consumption, $c$ is
the amount of money to be used for current consumption, $m(t)$ is the
amount of money holding at time $t$ and $r$ is the interest rate
prevailing in the market. This is a standard utility maximization
problem and solving it by Lagrange multiplier,
we get the optimal allocation as
 $c^*=(1-\lambda) m(t)$ and $f^*=(1+r)\lambda m(t)$.

\item[(ii)] The $i-$th agent invests $(1-\lambda_i) m_i(t)$ in the market
and produces an output vector $y_i(t)$ 
which he sells in the market at some market determined price vector $p_t$ which
is same for everybody.
By the assumption of {\it perfect competition} we get that
$$(1-\lambda_i) m_i(t)= p(t) y_i(t).$$
The argument is roughly the following. If {\it l.h.s} $\geq ${\it r.h.s}, then it
is not optimal to produce because cost is higher than revenue. On the other hand,
if {\it r.h.s} $\geq$ {\it l.h.s}, then there exists what is called {\it supernormal}
profit which attracts more agents to produce more. But that leads to a fall in price
and hence the economy comes to the equilibrium only when {\it l.h.s} $=$ {\it r.h.s}.
Summing up the above equation over all agents, we get
$$\Sigma_i (1-\lambda_i)m_i(t)= p(t)\Sigma_i y_i(t).$$
\noindent The above equation can be rewritten as 
$$M(t)V(t)=p(t) Y(t),   \eqno(8)$$
\noindent where $M(t)$ is the total money in the system and $V(t)$ is equivalent to 
the velocity of money at time $t$. Clearly,
$V(t)$ depends on the parameter of the utility functions $\lambda_i$ for all agents. 
It may be noted that the derived equation is analogous to the Fisher
equation of `quantity theory
of money' \cite{mankiw}. For an alternative interpretation of the Fisher equation
in the context of CC type exchange models, see ref \cite{ning ding}.

\item[(iii)] We have considered a closed economy. During the exchange process
money is neither created
nor destroyed. After all trading are done, each agent has whatever they
saved for future consumption and the interest income earned from it
added to some fraction $\alpha_i(t)$ of the total
amount of money invested in production of current consumption.
Briefly
$$m_i(t+1)=(1+r)\lambda_i m_i(t)+\alpha_i(t)\Sigma_i (1-\lambda_i) m_i(t)$$
\noindent or from eqn. (8),
$$m_i(t+1)=(1+r)\lambda_i m_i(t)+\alpha_i(t)p(t) Y(t). \eqno (9)$$
\noindent Let us assume $r=0 $ and $\alpha_i(t)p(t) Y(t)=\epsilon(t)$. We get the following
reduced equation,
$$m_i(t+1)=\lambda_i m_i(t)+\epsilon(t). \eqno(10)$$
\end{enumerate}
\subsection{Steady state distribution of money and price}
\noindent 
The money exchange process
of each agent is governed by eqn. $(10)$ where $\epsilon(t)$ can be assumed to be 
{\it white noise}. Then it has been shown that
this process produces the gamma function-like part as well as
the power-law tail \cite{basu}. We now consider the more general
version of it viz. eqn. $(9)$. 
Clearly this is an autoregressive process of order $1$ with $(1+r)\lambda_i<1$ 
assuming that the last 
term is white noise.
If we take
expectation over the whole expression we get
$$[1-(1+r)\lambda_i]\langle m_i\rangle =\langle\alpha_i(t)\rangle \langle p(t)\rangle
\langle Y(t)\rangle.$$
\noindent We rewrite the equation in terms of average money holding (without subscript) 
and denoting $\langle\alpha_i(t)\rangle\langle p(t)\rangle\langle Y(t)\rangle$
by a finite constant $C$, as
$$\lambda=\frac{1}{1+r}(1-\frac{C}{m})$$
\noindent which implies $d\lambda\propto dm/m^2$.
\noindent Since $P(m)dm= \rho(\lambda)d\lambda$ where $P(m)$ is the distribution of money
and $\rho(\lambda)$ is the distribution of $\lambda$, we have
$$P(m)=\rho\left(\frac{1}{1+r}(1-\frac{C}{m})\right)\frac{1}{m^2}.  \eqno(11)$$
\noindent For example if $\lambda$ is distributed uniformly then the distribution
of money has a power law feature with the exponent being $2$. Similar
argument is present in refs. \cite{chatterjee, basu}. Ref. \cite{basu} also presents
examples of the emergence of gamma function-like behavior in the
distribution of money for various types of noise terms.

\noindent We now focus on eqn. $(8)$. We rewrite it without subscript
as following.
$$p = \frac{V}{Y/M}$$
Note that $M$ is of the order of $N$, the number of agents whose total production
is $Y$. In short run fluctuations in output takes place for several reasons.
We assume that both $V$ and $Y/M$ are distributed uniformly. It may
be shown in that case that, the distribution of price is a power law,
$$f(p)\sim p^{-2}.  \eqno(12)$$
Hence in this model price also may have a power law fluctuation. It may be noted that
there is no clear evidence supporting the existence of a power law
in commodity price fluctuation. But it has been verified in stock price
fluctuations (see e.g. ref. \cite{sornette}).

\section{Summary and discussion}

\noindent 
Our primary focus was to develop a minimal microeconomic framework to derive the asset
equations used in the 
ideal gas like market models. We see that the framework considered above can very easily
reproduce the exchange equations used in the CC model (with fixed savings parameter).
In a certain limit, it also produces the exchange equations with complete
random sharing of monetary assets. 
Based on this model we have proposed an ideal gas like model of income distribution and
we have shown that it captures the gamma function-like behavior
of the real income distribution quite well. As discussed above, the framework
considered here and the resulting exchange equations differ significantly 
from those considered in \cite{lux,jet}. The utility function (in the basic microeconomic 
model considered above) deals with the behavior of the agents
in an exchange economy. However, it also captures the behavior of traders of put and call options 
of the same stock in a stock market. The price of call and put options of a particular stock generally 
vary inversely, depending on strike prices and expiration dates. An exception to this 
generalization is periods of symmetric volatility in the stock's price, when the simultaneous 
purchase of call and put options, a straddle, may be profitable. An 
option's price (particularly the log of  proportional return) is readily identified with its utility. Further, 
$\lambda$ may be slightly re-interpreted from determining the utility of savings to
determining the utility of protecting savings from risky trading. 
An interesting question would be whether the stationary distribution of the CC model returns in this model of
option trading or not. If not, what modification of the CC model might?

\noindent Next, we have analyzed the Monte Carlo simulation results by considering the
first two moments of the income distribution.  
The same has been done to analyze the income distributions 
produced by the CC model. Only the moment considerations
in both the models show the transition from exponential to delta function with 
changes in the parameter values of the respective models 
(the rate of transfer in case of the transfer model
and the rate of savings in case of CC model).
Moreover, the values of the income distribution parameters, conjectured in ref. \cite{kaski},
have been derived here only by considering those moments.
Next, the initial microeconomic model is generalized by incorporating an output market
(with global price determination) explicitly. The asset evolution equation in this
context is seen to be represented well by an autoregressive processe which very easily
produces a power law distribution of assets. 
It has already
been discussed in ref. \cite{basu} in details how such a process 
can generate
insightful results regarding the distribution of monetary assets.  
In the same context, an aggregative equation is derived which is analogous to
the Fisher equation. From this equation,
a power law in price fluctuation is also derived. 
Taken together, these models provide a link between the standard microeconomic
settings (individual optimization and output market) and
the asset exchange equations used in the ideal gas like market models.

\vskip 1 cm
\noindent {\bf Acknowledgement}: We are thankful to an anonymous referee for suggesting several 
useful insights and technicalities, commonly accepted in standard economic literature.

\section{Appendix}
{
{\bf A1:} Demand functions derived from the utility maximization problem
are as follows. For agent 1,
$$x^*_1=\alpha_1\frac{(M_1+p_1Q_1)}{p_1},~~~ x^*_2=\alpha_2\frac{(M_1+p_1Q_1)}{p_2},~~~
m^*_1=\alpha_m(M_1+p_1Q_1).$$
\noindent Similarly for agent 2,
$$y^*_1=\alpha_1\frac{(M_2+p_2Q_2)}{p_1},~~~y^*_2=\alpha_2\frac{(M_2+p_2Q_2)}{p_2},~~~
m^*_2=\alpha_m(M_2+p_2Q_2).$$
\noindent Now, we equate demand and supply of both commodities (i.e. 
$x^*_1+y^*_1=Q_1$ and $x^*_2+y^*_2=Q_2$).
By substituting the values of $x^*_1$, $x^*_2$, $y^*_1$ and $y^*_2$ and
by solving these two equations we get market clearing prices 
($\hat p_1$, $\hat p_2$) where
$$\hat p_1=\frac{\alpha_1}{\alpha_m}\frac{(M_1+M_2)}{Q_1}~~~and~~~
\hat p_2=\frac{\alpha_2}{\alpha_m}\frac{(M_1+M_2)}{Q_2}.$$
{\bf A2:} Consider the demand functions for money (i.e. $m^*_1$, $m^*_2$)
at optimal prices ($\hat p_1$, $\hat p_2$). Clearly,
$$m^*_1+m^*_2=\alpha_m(M_1+M_2)+\alpha_m(\hat p_1Q_1+\hat p_2Q_2).$$
Substituting the value of ($\hat p_1,\hat p_2$) in the above equation and using 
that $(\alpha_1+\alpha_2+\alpha_m)=1$, we get the desired result that
$$m^*_1+m^*_2=(M_1+M_2).$$
{\bf A3:} From {\bf A1} we get in equilibrium,
$$m^*_1=\alpha_m(M_1+\hat p_1.Q_1).$$
\noindent Substituting the value of $\hat p_1$ in the above equation we get
$m^*_1=\alpha_mM_1+\alpha_1(M_1+M_2)$ which can be written as
$$m^*_1=\alpha_mM_1+\frac{\alpha_1}{\alpha_1+\alpha_2}(1-\alpha_m)(M_1+M_2).$$ 
\noindent For agent $2$ the corresponding equation is
$$m^*_2=\alpha_mM_2+\frac{\alpha_2}{\alpha_1+\alpha_2}(1-\alpha_m)(M_1+M_2).$$ 
Denoting  $\alpha_m$ by $\lambda$ and $\alpha_1/(\alpha_1+\alpha_2)$ by $\epsilon$, we get the desired asset equations of CC model. Since the parameters add up
to 1, if $\alpha_1$ is uniform over [0,1-$\lambda$] then so is $\alpha_2$.

{\bf A4:} For the $i-$th agent, the rule of trading is the following
$$m_i(t+1)=g[m_i(t)+m_j(t)].$$
Applying expectation operator on both sides, we get
$$\langle m_i\rangle = \langle g\rangle[\langle m_i\rangle+
\langle\frac{1}{N}\Sigma_j m_j\rangle].$$
Writing the above equation without subscript and using the fact that $\langle g\rangle
=\frac{1}{2}$, 
we get
$\langle m \rangle =\frac{1}{2}[\langle m\rangle+1]$. This in turn gives $\langle m \rangle = 1$.

Also, in the steady state
$\Delta m_i=\Delta[g(m_i+m_j)]
=\langle x^2\rangle-(\langle x\rangle)^2$, where $x=[g(m_i+m_j)].$
\noindent Note that $\langle x\rangle=1$. Hence
$$\Delta m=\langle g^2\rangle\langle m^{2}_i+m^{2}_j+2m_im_j\rangle-1.$$
\noindent Using the fact that $m_i$ and $m_j$ are uncorrelated and $\Delta g=
\langle g^2\rangle-1/4$, we get
$$\Delta m=(\Delta g+\frac{1}{4})(2\Delta m+4)-1.$$
\noindent Simplifying, we get
$$\Delta m=\frac{4\Delta g}{\frac{1}{2}-2\Delta g}.$$
{\bf A5:} 
\noindent For the $i$-th agent, the rule of trading is the following
$$m_i(t+1)=\lambda m_i(t)+\epsilon(1-\lambda)[m_i(t)+m_j(t)].$$
\noindent Following {\bf A4}, by taking expectations on both sides we get
$\langle m\rangle = 1$. Also, in the steady state,
$\Delta m_i=\langle x^2\rangle-(\langle x\rangle)^2$ 
where $x=\lambda m_i+\epsilon(1-\lambda)(m_i+m_j)$.

\noindent Using the fact that $\langle x\rangle = 1$, we get
$$\Delta m_i=\lambda^2 \langle m^{2}_i\rangle+(1-\lambda)^2\langle \epsilon^2(m_i+m_j)^2\rangle+2
\lambda(1-\lambda)\langle \epsilon\rangle\langle m_i(m_i+m_j)\rangle-1.$$
\noindent Using the result from {\bf A4} that $\langle \epsilon^2(m_i+m_j)^2\rangle=
(\Delta \epsilon+1/4)(2\Delta m+4)$ and the result that $\Delta \epsilon=1/12$, we get
the following equation after rearranging terms
$$\Delta m=\lambda^2(\Delta m+1)+\frac{2}{3}(1-\lambda)^2(\Delta m+2)+\lambda(1-\lambda)
(\Delta m+2)-1.$$
\noindent Simplifying the above expression we get the desired result,
$$\Delta m=\frac{(1-\lambda)^2}{(1-\lambda)(1+2\lambda)}.$$
}

\section{Appendix - annex}
{
\noindent This section is for record and an appeal; not for publication. We had the rare fortune 
to have an economist as a referee who has been extremely supportive of such 
econophysics research, unusually encouraging (see the excerpts below), very caring yet critical 
(more than 3 pages of referee reports; even checking some of the results independently), suggesting
several improvements to make the presentation more acceptable to the economists as well. There 
have also been some suggestions for future research to establish the points.

\medskip

\noindent The introduction of the referee report reads:

\medskip

\noindent ``{\it Stochastic particle system models in which particles exchange a positive quantity  
that is conserved (sum of quantity over all particles remaining constant) have been shown to 
have stationary distributions that resemble in some ways distributions of personal labor income 
and related income measures (e.g., household income). A challenge facing researchers interested in 
particle systems is to synthesize these findings with conventional economics. Working in the 
opposite direction (economic theory to a plausible function for income distribution) have been 
generations of economists. Since Pareto's time, economists have struggled to variously derive 
distributions of labor income, household income, personal asset income, small business net income, 
and large business net income  from neoclassical micro-economics. These efforts have succeeded in 
plausibly explaining the mean of such distributions. Explaining the second moment is a work in progress. 
The most successful effort - at
least the most widely accepted as indicated by the frequency of its appearance in introductory 
textbooks - is the derivation of the log-normal distribution as a model of personal income. 
This derivation has a well known problem with its second moment but the simplicity of the stochastic 
generator is nevertheless appealing.}"

\noindent ``{\it{\bf Finding how micro-economic theory implies the distribution of labor income is what the ascent 
of the tallest peak of the Himalayas was to mountain climbers before Tenzing Norgay stepped on 
to its summit.  [This paper] attempt this feat. If it succeeds, it is a landmark paper.} Just 
making progress toward its objective, just ascending to a `col' no one else has yet gotten to, 
is a notable paper deserving publication. {\bf A number of economists who later won the Nobel Prize 
in economics attempted to solve the problem of deriving a function for labor income distribution, 
when they were young and ambitious.} Their efforts were published. No one thinks that anyone 
has plausibly solved the problem yet. Until recently economists did not consider particle system 
models of income distribution.}"

\noindent ``{\it... [this] paper represents a 
clever synthesis of conservative particle system model and familiar micro-economic concepts 
and equations. Such a synthesis, even an empirically implausible one, is important. ... 
}"

\medskip

\noindent The report further mentions:

\medskip

\noindent ``{\it ... The paper's integration of basic equations of mathematical economics with 
the CC model is interesting and worth publishing. The paper's implausibilities as economics 
will be apparent to economists and will stimulate tinkering with the paper's model. ...
I [had the] advantage from having seen the paper before publication.}"

\medskip

\noindent We are aware that our effort here is modest and requires further developments.
We would like to appeal to the economists, and the referee in particular, to join
the effort in establishing the ideal gas like market models (e.g., CC-CCM models) with
microeconomic foundations.
This derivation of the income distribution from microeconomic theory is, as the referee mentioned, long overdue.

}

\end{document}